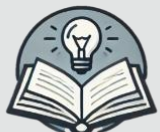

# Do more citations mean better patents?

Gaétan de Rassenfosse

Holder of the Chair of Science, Technology, and Innovation Policy
École polytechnique fédérale de Lausanne, Switzerland.

This version: August 2025

**Purpose**
This article is part of a Living Literature Review exploring topics related to intellectual property, focusing on insights from the economic literature. Our aim is to provide a clear and non-technical introduction to patent rights, making them accessible to graduate students, legal scholars and practitioners, policymakers, and anyone curious about the subject.

**Funding**
This project is made possible through a Living Literature Review grant generously provided by Open Philanthropy. Open Philanthropy does not exert editorial control over this work, and the views expressed here do not necessarily reflect those of Open Philanthropy.

## Do more citations mean better patents?

Gaétan de Rassenfosse
École polytechnique fédérale de Lausanne, Switzerland.

Patent documents cite earlier patents, much like scientific articles reference prior research. Scholars and practitioners routinely use the number of citations a patent receives as a proxy for invention importance, based on the assumption that patents attracting more citations are more valuable. This article reviews what the evidence says about that assumption.

**An intuitive idea, but…**

The intuition is simple and powerful: valuable ideas attract attention. Nobel Prize laureates, for instance, almost invariably have highly cited papers to their name. Citation counts, in this view, provide a tidy signal of impact.

However, while scientific citations are widely understood as a way to give credit—recognizing a researcher's contribution to the advancement of knowledge—patent citations serve a different purpose, rooted in legal procedure. Citations in patents are part of the legal assessment of patentability, disclosing prior art that helps examiners determine whether an invention is novel and non-obvious. This legal function blurs the use of patent citations as signals of value—at least in the way we interpret citations in science.

And then there's the question of value itself, as discussed in an article on patent quality.[1] What exactly do we mean by a "valuable" patent? Are we talking about the value of the underlying invention, or the legal right it confers? Are we referring to technological value or economic value? And if economic, is it private value to the firm, or social value to the world?

**Not all patent citations are the same**

Before diving into the empirical literature, a word of caution about patent citations is warranted. Virtually all studies rely on so-called front-page citations—the references printed on the first page of granted patent documents. These citations are standardized and widely available, making them a convenient data source.

However, the patenting process gives rise to other types of citations. Depending on the jurisdiction, these include references listed in <u>Information Disclosure Statements</u> (IDS) submitted by applicants, prior art discussed in <u>Office actions</u> issued by examiners, and citations that may appear in the body of the patent text. Additional citations may also emerge in post-grant proceedings, such as oppositions. These alternative sources are less frequently used in research but may offer different insights into how patents engage with prior art.

**The first validation study**

We owe the first validation study of patent citations as a signal of value to information scientist Francis Narin and colleagues at CHI Research, one of the earliest firms in patent

---

analytics. Observing the use of citation-based metrics in science, they set out to test whether patent citations could similarly be used to identify important patents (Carpenter et al. 1981).

Their approach was elegantly simple. They identified patents behind products that received the IR-100 award between 1968 and 1974. This award, granted by the *Industrial Research and Development* magazine, recognizes the most significant new technical products of the year. For comparison, they selected a set of 100 "control" patents randomly drawn to match the time distribution of the award-winning group. The result? The award-related patents were cited more than twice as often as the controls. The authors concluded that "citation analysis will be useful in identifying important patents."

However, the study was vague about what exactly was being measured. The IR-100 award recognized "products of certified industrial or commercial importance," which may reflect technical merit, economic potential, or a combination of both.

**Citations and the "technical merit" of the invention**

A decade later, a follow-up study by the same group looked specifically at technological importance (Albert et al. 1991). The authors examined 77 patents related to silver halide technology granted to Eastman Kodak between 1982 and 1983. They asked 20 senior Kodak scientists to rate how much each patent had advanced the state of the art in the field. The results showed a significant relationship between expert ratings and citation counts.

Interestingly, the signal was strongest at the top of the distribution: highly cited patents were more likely to be rated as "important." At the lower end—between patents with zero or only a few citations—there was little discernible difference in expert ratings.

**A handful of other studies**

Only a handful of additional validation studies have examined patent citations as a proxy for the technological importance of an invention.

One recent example is a study by Capponi et al. (2022), which follows the tradition of linking citations to external markers of innovation quality. The authors focus on patents protecting products that received a UK Queen's Award for Innovation. Specifically, they identified 1,468 patent families filed between 1976 and 2013, covering 401 award-winning innovations. The number of citations these patents received was a strong predictor of award status.

Another notable example comes from Czarnitzki et al. (2011), who also rely on expert judgment—but with a twist. They focus on 188 "wacky" patents, collected by an employee of the World Intellectual Property Organization. These patents were selected for their futile or absurd nature, typically involving only a marginal inventive step or barely satisfying the non-obviousness criterion. The authors then compared the citation rates of the wacky patents to those of a control group filed in the same year and technology class. The result: wacky patents received significantly fewer citations.

Finally, Moser et al. (2018) adopted a distinct approach. Their study links patents to objective measures of improvement in the quality of the patented invention, using hybrid corn as their empirical setting. Unlike studies based on expert opinion, improvements in agricultural performance can be directly observed in field trial data, by comparing a patented hybrid's yield to that of the best-performing comparison hybrids. They analyzed 315 patent–hybrid corn pairs between 1985 and 2002 and found that citations strongly correlate with improvements in yield, although this relationship was driven primarily by applicant-added citations.

**Citations and the "economic merit" of the invention**

Trajtenberg (1990) offered a different kind of validation—one grounded in economics. Using the rise of Computed Tomography (CT) scanners in the U.S. market between 1973 and 1982, he documented a strong link between citation-weighted patent counts and the social value of innovation, measured in terms of both consumer surplus and producer profits.

The intuition behind his approach is as follows: imagine that all available CT scanners can be described using a handful of key attributes, say, resolution, scan time, and safety features. Over time, innovation leads to the introduction of new scanner models and improvements in existing ones. These product changes expand consumer choice and generate value, either by improving quality or lowering prices. Using tools from industrial organization, Trajtenberg quantifies that value based on detailed hospital purchase data and scanner characteristics. He then shows that yearly improvements in social value metrics correlate with a citation-based count of CT scanner patents. While the statistical analysis is impressive, the reliance on yearly aggregated data means that the evidence ultimately rests on just ten data points, calling for more validation studies.

A flurry of studies has since examined the economic value of patents. These studies draw on a range of data sources to estimate value. Still, their focus is primarily on the private economic value of an invention—that is, the monetary benefits that an invention generates for its owner.

**Survey-based measures of value**

One group of studies has collected direct, subjective estimates of patent value from inventors or patent-holding firms through surveys.

Harhoff et al. (1999) regressed the number of citations received by 192 U.S. patents filed in the late 1970s on patent owners' retrospective estimates of private value. The key survey question asked inventors to estimate how much their patent would have been worth—in hindsight—knowing what they now knew about the profits the invention ultimately generated. The study finds a significant positive correlation between citation counts and these retrospective value estimates.

A similar approach was used in a large-scale survey of inventors by Gambardella et al. (2008). Drawing on responses from the PatVal-EU survey—covering more than 9,000 European patents with priority dates from 1993 to 1997—the authors tested the relationship between citations and inventor-assessed value. They find that citations are the strongest predictor of value among several indicators (including patent family size). However, their statistical model explains only 2.7% of the variation in private value. This low number suggests that while citations carry some signal, they are a noisy indicator of value.

**Exploiting auction price data**

While survey-based measures offer a rare window into how inventors themselves assess patent value, they suffer from important limitations—including recall bias, selective responses, and inconsistent interpretations. These concerns have motivated researchers to explore alternative, more objective measures.

One such effort comes from Sneed and Johnson (2009), who investigated the determinants of patent value in an auction setting. Auction prices provide a directly observed measure of private economic value, based on what buyers are actually willing to pay.

The authors analyzed a dataset of 402 patents grouped into 99 lots from Ocean Tomo's live patent auctions held in 2006, including 49 sold and 50 unsold lots. They found a positive relationship between citation counts and auction value. Each additional citation was associated with approximately $10,600 in additional price per patent. That said, patent auctions remain a relatively thin and selective market—not all patents are auctioned, and those that are may differ systematically from the broader population. As such, while auction prices offer a cleaner measure of value than surveys, they come with their own selection and generalizability concerns.

Note that data from Ocean Tomo's auctions have also been used by Nair et al. (2011), Fischer and Leidinger (2014), and Odasso et al. (2015). These studies systematically find a positive correlation between the auction price and the count of citations.

**Stock market data**

Another approach to estimating patent value draws on stock market data. A classic study in this stream is that of Hall et al. (2005), who exploit Tobin's q, a widely used measure in finance that compares a firm's market value to the replacement cost of its assets (that is, what it would cost to rebuild the firm from scratch). A high Tobin's q suggests that investors believe the firm owns valuable intangible assets—like knowledge, brand, or intellectual property.

The authors link more than half a million U.S. patents to publicly-traded firms, constructing three measures of firms' "knowledge stock": R&D spending, patent counts, and citation-weighted patents. They find that all three indicators explain variation in Tobin's q, but citations per patent turn out to be particularly informative. Firms with more highly cited patents tend to be valued more highly by the market, even after accounting for their R&D spending and number of patents. In other words, citations help distinguish high-value patents from run-of-the-mill ones, and this distinction shows up in firms' stock prices.

Kogan et al. (2017) offer a more direct use of stock price data. Their method estimates the private economic value of individual patents by analyzing how the market reacts in a narrow window around each patent's grant date. They find a strong and positive relationship between this market-based value estimate and citation counts.

A note of caution is warranted, however. While stock market data offer a powerful lens for assessing patent value, they come with limitations. Most importantly, stock prices reflect investor expectations, which may be influenced by hype, noise, or factors unrelated to the patent's actual economic returns. Furthermore, this assessment occurs early in the patent's life, and the value may not be realized.

**Patent renewal data**

Another empirical strategy for validating patent citations as a proxy for private value draws on patent renewal records. Maintaining a patent over its full term requires periodic fee payments, and inventors are expected to make these payments only when expected benefits outweigh the costs. Renewal behavior thus offers a revealed-preference indicator of a patent's private value. Bessen (2008) applies this method to estimate the value of U.S. patents and finds that highly cited patents tend to be more valuable.

Renewal-based value estimates capture only the lower bound of expected value—enough to justify paying the fee—but not the full distribution of returns. Moreover, some patents may be renewed for strategic reasons unrelated to immediate commercial value, such as blocking competitors or signaling capability.

**Administrative records**

Finally, Giummo (2010) explores a unique source of data for patent valuation: inventor compensation records mandated by the German Employee Invention Act of 1957. This law requires firms to compensate employees for patented inventions they create in the course of their work, based on the economic benefit derived from the invention's use.

Using these records, Giummo assembles a dataset of over 1,100 U.S. patents of German origin filed between 1977 and 1982. He finds a positive correlation between citation counts and inventor compensation. This finding supports the idea that citations capture the economic relevance of patented inventions, even when value is measured in terms of use. The study thus adds further evidence that citation data reflect meaningful differences in patent value.

**Concluding remarks**

The evidence reviewed here suggests that patent citations can serve as a meaningful proxy for patent importance. Their appeal lies in their accessibility and scalability: they are publicly available, quantifiable, and cover the entire patent universe. Yet, as with any proxy, they should be used with care.

Validation studies span three conceptions of value: technical merit, private economic value, and social value. The evidence is uneven across these categories. For technical merit, the number of studies is limited and often based on small expert-rated samples, yet they consistently find a positive correlation with citations. For private economic value, the literature is more extensive and methodologically diverse, consistently finding that citations correlate with private value. However, these correlations explain only a modest share of the overall variation in patent value. For social value—the contribution of inventions to societal welfare—there is only one major study, highlighting a clear need for more research.